\documentclass[showpacs,aps,prd,,nofootinbib,floats]{revtex4}
\usepackage{graphicx}
\usepackage{amssymb,latexsym}
\usepackage[draft=false]{hyperref}

\def\lsim{\hbox{ \raise.35ex\rlap{$<$}\lower.6ex\hbox{$\sim$}\ }}
\def\gsim{\hbox{ \raise.35ex\rlap{$>$}\lower.6ex\hbox{$\sim$}\ }}
\def\setC{\mathbb{C}}
\def\setR{\mathbb{R}}

\begin{document}

\title{${D}$-term inflation, cosmic strings, and consistency with cosmic
    microwave background measurements}

\author{Jonathan Rocher}
\email{rocher@iap.fr}
\affiliation{Institut d'Astrophysique
de Paris, {${\cal G}\setR\varepsilon\setC{\cal O}$}, FRE 2435-CNRS, 98bis
boulevard Arago, 75014 Paris, France.}

\author{Mairi Sakellariadou} \email{msakel@cc.uoa.gr, mairi@mpej.unige.ch}
\affiliation{Division of Astrophysics, Astronomy, and Mechanics, Department of
Physics, University of Athens, Panepistimiopolis, GR-15784 Zografos, Hellas,
and \\ D\'epartement de Physique Th\'eorique, Universit\'e de Gen\`eve, 24
quai E. Ansermet, CH-1211 Gen\`eve 4, Switzerland.}

                   
\begin{abstract} 
Standard D-term inflation is studied in the framework of
supergravity. D-term inflation produces cosmic strings,
however it can still be compatible with CMB measurements without
invoking any new physics.  The cosmic strings contribution to the CMB
data is not constant, nor dominant, contrary to some previous
results. Using current CMB measurements, the free parameters (gauge
and superpotential couplings, as well as the Fayet-Iliopoulos term) of 
D-term inflation are constrained.
\end{abstract}

\maketitle

\section{Introduction}

The inflationary paradigm~\cite{infl} offers simple answers to the
shortcomings of the standard hot big bang model. In addition, simple
inflationary models offer successful candidates for the initial
density fluctuations leading to the observed structure formation.  One
crucial question though is to answer how generic is the onset of
inflation~\cite{onset} and to find consistent and natural models of
inflation from the point of view of particle physics.  One can argue
that the initial conditions which favor successful inflationary models
are the likely outcome of the quantum era before
inflation~\cite{onset}. It is more difficult however to find
natural ways to guarantee the flatness  of the inflaton potential.

The early history of the Universe at energies below the Planck scale
is described by an effective N=1 supergravity theory. Since inflation
should have taken place at an energy scale $V^{1/4}\lesssim 4\times
10^{16}$ GeV, this implies that inflationary models should be
constructed in the framework of supergravity. Here is where the
problem arises: it is difficult to implement slow-roll inflation
within supergravity. The positive false vacuum of the inflaton field
breaks spontaneously global supersymmetry, which gets restored after
the end of inflation (when $V$ disappears). In supergravity theories,
the supersymmetry breaking is transmitted to all fields by gravity,
and thus any scalar field, which could play the r\^ole of the
inflaton, gets an effective mass $\sim\sqrt{8\pi V}/M_{\rm Pl}\sim H$,
where $H$ stands for the expansion rate during inflation, and $M_{\rm Pl}$
denotes the reduced Planck mass. This
problem, known as the problem of ``Hubble-induced mass'', originates
from F-term interactions and thus it is resolved if we consider the
vacuum energy as being dominated by non-zero D-terms of some
superfields~\cite{dterm}. This result led to a dramatic interest in
D-term inflation, since in addition this model can easily be implemented 
in string theory. 
However, later on D-term inflation in its turn was
though to be plagued with problems.

In D-term inflation, the inflationary era ends when a U(1)
gauge symmetry is spontaneously broken by a vacuum expectation value
of some scalar field, leading to the formation of gauge cosmic strings.  
As it was explicitly shown in Ref.~\cite{rjs}, cosmic strings are
generically expected to be formed at the end of a hybrid inflation phase, 
in the context of supersymmetric grand unified theories. 
It was claimed~\cite{rj} that the cosmic strings
contribution to the angular power spectrum of the Cosmic Microwave
Background (CMB) temperature anisotropies is constant and dominant
($75\%$). From the observational point of view however, strong
constraints~\cite{bprs} are placed on the allowed cosmic strings
contribution to the CMB: it can not exceed $\sim 10\%$. 
Thus, standard D-term inflation was thought
to be inconsistent with cosmology.  To rescue D-term inflation there
have been proposed different mechanisms which either consider more
complicated models, or they require additional ingredients so that
cosmic strings are not produced at the end of hybrid inflation. For
example~\cite{nocs}, one can add a nonrenormalisable term in the
potential, or add an additional discrete symmetry, or consider GUT
models based on non-simple groups. More recently, a new pair of charged
superfields has been introduced in D-term inflation so that cosmic
strings formation is avoided~\cite{jaa}.

The aim of our study is to show that standard D-term inflation
leading to the production of cosmic strings is still compatible with
cosmological data, and in particular CMB, without invoking any new
physical mechanisms. We find that in D-term inflation the cosmic
strings contribution to the CMB data depends on the free parameters, as for
F-term inflation~\cite{rs}.  The maximum allowed cosmic
strings contribution to the CMB measurements places upper limits on
the inflationary scale (which is also the cosmic string energy scale),
or equivalently on the coupling of the superpotential.

We first review the results for F-term hybrid inflation, in which case
the supersymmetric renormalisable superpotential reads
\begin{equation}\label{superpot}
W_{\rm infl}^{\rm F}=\kappa S(\Phi_+\Phi_- - M^2)~,
\end{equation}
where $S, \Phi_+, \Phi_-$ are three chiral superfields, and $\kappa$,
$M$ are two constants.  The cosmic strings contribution to the CMB is
a function of the coupling $\kappa$, or equivalently of the mass scale
$M$.  It can be consistent
with the most recent measurements, which require that it is at most
equal to $9\%$~\cite{bprs}, provided~\cite{rs}
\begin{equation}
M\lsim 2\times 10^{15} {\rm GeV} ~~\Leftrightarrow ~~\kappa \lsim
7\times10^{-7}~.
\end{equation}
The above limit was obtained in the context of SO(10) gauge group.
Upper limits of the same order of magnitude are found for other gauge 
groups~\cite{rs}.

This result implies that F-term inflation leading to the production of
cosmic strings of the GUT scale can be compatible with measurements,
provided the coupling is sufficiently small. Thus, hybrid supersymmetric
inflation losses some of its appeal since it is required some amount
of fine tuning of its free parameter, $\kappa$ should be of the
order of $10^{-6}$ or smaller.  This constraint on $\kappa$ is in
agreement with the one given in Ref.~\cite{kl}.  The parameter
$\kappa$ is also subject to the gravitino constraint which imposes an
upper limit to the reheating temperature, to avoid gravitino
overproduction. Within supersymmetric GUTs, and assuming a see-saw mechanism to
give rise to massive neutrinos, the inflaton field will decay during
reheating into pairs of right-handed neutrinos.  Using the constraints
on the see-saw mechanism it is possible~\cite{SenoSha,rs} to convert
the constraint on the reheating temperature to a constraint on the
coupling parameter $\kappa$, namely $\kappa \lesssim 8\times
10^{-3}~$, which is clearly a weaker constraint.

The superpotential coupling $\kappa$ is allowed to get higher values,
namely it can approach the upper limit permitted by the gravitino
constraint, if one employs the curvaton
mechanism~\cite{lw2002}.  Such a mechanism can be easily
accommodated within supersymmetric theories, where one expects to have
a number of scalar fields.  For fixed $\kappa$, the cosmic strings
contribution decreases rapidly as the initial value of the curvaton
field, ${\cal\psi}_{\rm init}$, decreases. Thus, the WMAP measurements
lead to an upper limit on ${\cal\psi}_{\rm init}$, namely
${\cal\psi}_{\rm init}\lsim 5\times 10^{13}(\kappa/10^{-2})$ GeV~\cite{rs}.
This limit holds for $\kappa$ in the range $[5\times 10^{-5}, 1]$; for
lower values of $\kappa$, the cosmic strings contribution is always
suppressed and thus lower than the WMAP limit.

The above results hold also if one includes supergravity corrections.
This is expected since the  value of the inflaton 
field is several orders of magnitude below the Planck scale.

\section{D-term inflation}
D-term inflation is derived from the superpotential
\begin{equation}
\label{superpotD}
W^{\rm D}_{\rm infl}=\lambda S \Phi_+\Phi_-~,
\end{equation}
where $S, \Phi_-, \Phi_+$ are three chiral superfields and $\lambda$
is the superpotential coupling. D-term inflation requires the
existence of a nonzero Fayet-Illiopoulos term $\xi$, permitted only if
an extra U(1) symmetry beyond the GUT framework, is introduced. In the 
context of supersymmetry we calculate the radiative corrections leading to the
effective potential,
\begin{equation}
\label{VexactD}
V^{{\rm D}-{\rm SUSY}}_{\rm eff}(|S|) =
\frac{g^2\xi^2}{2}\left\{1+\frac{g^2}{16\pi^2}
\left[2\ln\frac{|S|^2\lambda^2}{\Lambda^2}+
(z+1)^2\ln(1+z^{-1})+(z-1)^2\ln(1-z^{-1})\right]\right\}~,
\end{equation}
where $z=\lambda^2 |S|^2/(g^2\xi)$, with $g$ the gauge
coupling of the U(1) symmetry and $\xi$ the Fayet-Illiopoulos term,
chosen to be positive; $\Lambda$ stands for a renormalisation scale.  
In the absence of the curvaton mechanism, the
quadrupole anisotropy is the sum of the inflaton field (scalar and tensor 
parts) and cosmic strings contributions and we normalise it to the COBE data.

We compute the mass scale of the symmetry breaking, given by $\sqrt\xi$, 
and we find that it increases with $\lambda$.  We then 
calculate the cosmic strings contribution to the temperature anisotropies.
We find that within supersymmetry D-term inflation is consistent
with CMB data provided the superpotential coupling $\lambda$ is quite
small, namely $\lambda\lesssim 3\times 10^{-5}$.

However, the dependence of $z_{\rm Q}$ (the index Q denotes the scale
corresponding to the quadrupole anisotropy) on the superpotential coupling
$\lambda$ results to values of the inflaton field $S_{\rm Q}$ above the Planck
mass.  This implies that the correct analysis has to be done in the framework
of supergravity.  For small values of the gauge coupling $g$, the study in the
context of supergravity becomes just the analysis within supersymmetry.  Some
previous studies~\cite{rj,jap} found in the literature kept only the first
term of the radiative corrections. We find that it is necessary to perform the
analysis using the {\sl full} effective potential, which we calculated
for minimal supergravity. More precisely, using a minimal
K\"ahler potential  ($K=|\phi_-|^2+|\phi_+|^2+|S|^2$) and a minimal gauge
kinetic function ($f(\Phi_i)=1$), the scalar potential
reads~\cite{rs}
\begin{equation}
\label{VexactDsugra}
V^{\rm D-SUGRA}_{\rm eff} =
\frac{g^2\xi^2}{2}\left\{1+\frac{g^2}{16\pi^2}
\left[2\ln\frac{|S|^2\lambda^2}{\Lambda^2}\exp\left({|S|^2\over M_{\rm
Pl}^2}\right)+
(z+1)^2\ln(1+z^{-1})+(z-1)^2\ln(1-z^{-1})\right]\right\}~,
\end{equation}
where $z=[\lambda^2 |S|^2/(g^2\xi)]\exp(|S|^2/M_{\rm Pl}^2)$.   
The number of e-foldings is
\begin{equation}
N_{\rm Q}={2\pi^2\over g^2}\, \int_1^{z_{\rm Q}} 
\frac{{\rm W}(c\,{\tilde z})}
{{\tilde z}^2 f({\tilde z})[1+{\rm W}(c\,{\tilde z})]^2}\, {\rm d}{\tilde z}~,
\end{equation}
where ${\rm W}(x)$ denotes the ``W-Lambert function'' defined by
${\rm W}(x)\exp[{\rm W}(x)]=x$, and $c\equiv
(g^2 \xi)/(\lambda^2 M_{\rm Pl}^2)$. The number of e-foldings
$N_{\rm Q}$ is thus a function of $c$
and $z_{\rm Q}$, for $g$ fixed.  Setting $N_{\rm Q}=60$ we obtain 
a numerical relation between $c$ and $z_{\rm Q}$ which allows us
to construct a function $z_{\rm Q}(\xi)$ and  express the three
contributions to the CMB only as a function of $\xi$. The total $(\delta T/T)$
is given by
\begin{equation}\label{eqnumDsugra}
\left[\left({\delta T\over T}\right)_{\rm Q-tot}\right]^2 \sim
\left(\frac{\xi}{M_{\rm Pl}^2}\right)^2\left\{ \frac{\pi^2}{90g^2}z^{-2}_{\rm
Q}f^{-2}(z_{\rm Q}) \frac{{\rm W}(c\,z_{\rm Q})} {\left[1+{\rm W}(c\, 
z_{\rm Q})\right]^2} +\left(\frac{0.77
g}{8\sqrt{2}\pi}\right)^2 +
\left(\frac{9\pi}{4}\right)^2\right\}~,
\end{equation}
where the three contributions come from the scalar and tensor parts 
of the inflaton field, and the cosmic strings, respectively.
We normalise the {\sl l.h.s.} of
Eq.~(\ref{eqnumDsugra}) to the COBE data,  i.e.,
$\left(\delta T/ T\right)_{\rm Q}^{\rm COBE} \sim 6.3\times 10^{-6}$,
and we solve it numerically to obtain $\xi$, and thus, the three 
contributions for  given values of $g$ and $\lambda$. 

The cosmic strings contribution to the CMB data, is found to be an increasing 
function of the mass scale $\sqrt{\xi}$, as shown in Fig.~\ref{xifig} below.

\begin{figure}[htbp]
\includegraphics[scale=.5]{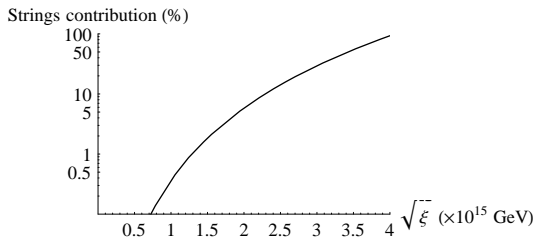}\\
\caption{The cosmic strings contribution to the CMB data, as a function of 
the mass scale $\sqrt{\xi}$ in units of $10^{15}$ GeV. }\label{xifig}
\end{figure}

\begin{figure}[htbp]
\includegraphics[scale=.45]{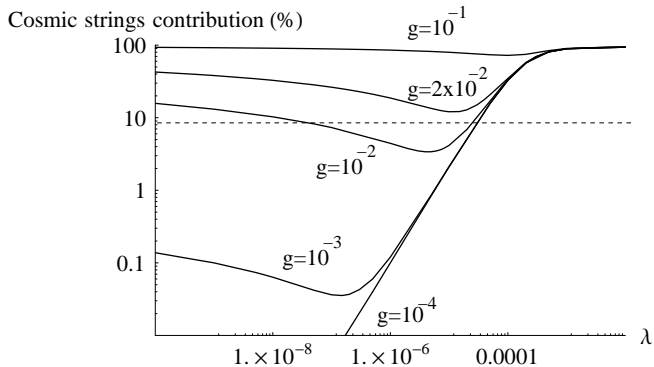}\\
\caption{Cosmic strings contribution to the CMB temperature anisotropies 
as a function 
of the superpotential coupling $\lambda$ for different values of the gauge 
coupling $g$. The maximal contribution allowed by WMAP is represented by a 
dotted line.}\label{PRL4}
\end{figure} 

Our results, summarized in Fig.~\ref{PRL4}, differ from the
results obtained in the framework of supersymmetry unless $\lambda
\gtrsim 10^{-3}$ or $g \lesssim 10^{-4}$. The cosmic strings
contribution to the CMB turns out to be dependent on the free parameters, 
with however the robust result that the cosmic strings contribution is not
constant, nor is it always dominant, in contradiction to
Ref.~\cite{rj}.  This implies that contrary to what is often assumed,
the simplest D-term inflation is still an open possibility and one
does not need to consider more complicated models. Our
analysis shows that $g\gtrsim 1$ necessitates multiple-stage
inflation, since otherwise we cannot have sufficient e-foldings to resolve 
the horizon problem of standard cosmology, while
$g\gtrsim 2\times 10^{-2}$ is incompatible with the WMAP measurements.  
For $g\lesssim 2\times
10^{-2}$, we can also constrain the superpotential coupling $\lambda$
and get $\lambda \lesssim 3\times 10^{-5}$. This limit was already
found in the framework of supersymmetry~\cite{rs} and it is in
agreement with the constraint $\lambda\lesssim {\cal O}(10^{-4}-10^{-5})$
of Ref.~\cite{jap}. 
Supergravity corrections impose in addition a lower limit to
the coupling $\lambda$. If for example $g=10^{-2}$, the cosmic strings
contribution imposes $10^{-8}\lesssim \lambda \lesssim 3\times
10^{-5}$.  The constraint induced by CMB measurements is
expressed as a single constraint on the Fayet-Iliopoulos term $\xi$,
namely $\sqrt\xi \lesssim 2\times 10^{15}~{\rm GeV}$.

\begin{figure}[htbp]
\includegraphics[scale=.5]{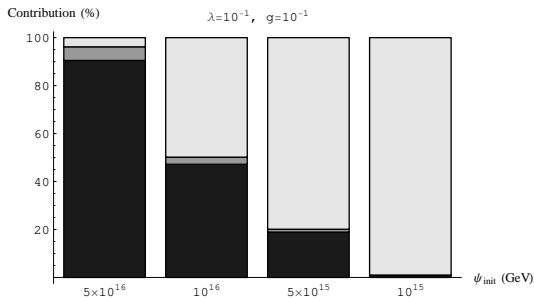}\\
\caption{The cosmic strings (dark gray), curvaton (light gray) and
inflaton (gray) contributions to the CMB temperature anisotropies as a
function of the the initial value of the curvaton field
${\cal\psi}_{\rm init}$, for $\lambda=10^{-1}$ and
$g=10^{-1}$.}\label{prl2}
\end{figure}

As a next step we examine whether there is a mechanism to 
allow more natural values of the couplings.  Assuming the existence of a scalar
field, that is subdominant during inflation as well as at the
beginning of the radiation dominated era, such a field (the curvaton)
gives an additional contribution to the temperature anisotropies,
which we calculate below for the case of D-term inflation.
The curvaton contribution, in terms of the metric perturbation, 
reads~\cite{mt2002}
\begin{equation}
\left(\frac{\delta T}{T}\right)_{\rm curv} \equiv \frac{\Psi_{\rm
    curv}}{3} ={4\over 9}{\delta{\cal \psi_{\rm init}}\over \psi_{\rm
    init}}~.
\end{equation}
The initial quantum fluctuations of the curvaton
    field are given by
$\delta{\cal \psi}_{\rm init}=H_{\rm inf}/( 2\pi)$.
The expansion rate during inflation, $H_{\rm infl}$, 
is a function of the inflaton field and it is given by the Friedmann 
equation: $H_{\rm infl}^2(\varphi)=(8\pi/3)V(\varphi)$.
Thus, for the D-term tree-level effective potential, 
the additional curvaton contribution to the total
temperature anisotropies is given by~\cite{jap}
\begin{equation}
\left[\left(\frac{\delta T}{T}\right)_{\rm curv}\right]^2={1\over
      6}\left({2\over 27\pi}\right)^2\left(
{g\xi\over M_{\rm Pl}\psi_{\rm init}}\right)^2~.
\end{equation}
Normalising the total $(\delta T/ T)$ to COBE we then obtain the
contributions of the different sources (inflaton field splitted into scalar 
and tensor parts, cosmic strings, curvaton field) to the CMB as a function 
of one of the three parameters $\psi_{\rm init}, \lambda, g$, keeping the 
other two fixed.
We show in Fig.~\ref{prl2} the three contributions as a function of
$\psi_{\rm init}$, for $\lambda=10^{-1}$ and $g=10^{-1}$.
Clearly, there are values of $\psi_{\rm init}$ which allow bigger
values of the superpotential coupling $\lambda$ and of the gauge coupling $g$,
than the upper bounds obtained in the absence of a curvaton field.

More explicitely, the fine tuning on the couplings can be avoided provided
\begin{equation}
\psi_{\rm init}\lsim 3\times 10^{14}\left({g\over 10^{-2}}\right) ~{\rm GeV}
~~~{\rm for}~~ \lambda\in [10^{-1}, 10^{-4}]~.
\end{equation}
Clearly, for smaller values of $\lambda$, the curvaton mechanism is not 
necessary.

We would like to bring to the attention of the reader that in the
above study we have neglected the quantum gravitational effects, which
would lead to a contribution to the effective potential, even though
$S_{\rm Q}\sim {\cal O}(10 M_{\rm Pl})$.  Our analysis is however
still valid, since the effective potential given in
Eq.~(\ref{VexactDsugra}) satisfies the conditions~\cite{lindebook}
$V(|S|)\ll M_{\rm Pl}^4$ and $m^2_S={\rm d}^2 V/{\rm d}S^2 \ll M_{\rm
  Pl}^2$, and thus the quantum gravitational corrections $[\Delta
V(|S|)]_{\rm QG}$ are negligible when compared to the effective
potential $V_{\rm eff}^{\rm D-SUGRA}$.

\section{Conclusions}
D-term inflation gained a lot of interest since it was shown that it 
avoids the problem of ``Hubble-induced mass'', but it was later thought 
to be plagued with an inconsistency with the data. 
Standard D-term inflation ends with the formation of cosmic
strings which was claimed to lead  to a constant and dominant contribution 
to the CMB data, much higher than the one allowed by measurements. 
In this study, we show that this is not the case, and therefore, 
standard D-term hybrid 
inflation can still be compatible with cosmological data.

We consider standard D-term inflation in its simplest form and without
any additional ingredients. We perform our analysis in the
framework of supergravity, since we reach scales above the Planck
scale, and we consider all one-loop radiative corrections. We show that 
the cosmic strings produced at the end of D-term inflation can lead to a 
contribution to the CMB data which is allowed by the measurements. 
The price to be paid is that the couplings must be small. 
However, this constraint can be less severe
if one invokes the curvaton mechanism.

\section*{Acknowledgements}
It is a pleasure to thank 
 G. Esposito-Far\`ese,  A.\ Linde and P.\ Peter
 for discussions and comments.

\end{document}